\documentclass[aps,pra, twocolumn]{revtex4-2}
\usepackage{amsthm, comment}
\usepackage[utf8]{inputenc}
\usepackage[sc,osf]{mathpazo}
\usepackage{graphicx,subfigure}
\usepackage{makeidx} 
\usepackage{amsmath}
\usepackage{amssymb}
 \usepackage{color, xcolor}
\usepackage{amsthm, MnSymbol}
\usepackage{hyperref}
\usepackage{cleveref}
\usepackage{diagbox}
\hypersetup{colorlinks, citecolor=blue, filecolor=blue, linkcolor=blue, urlcolor=blue}
\usepackage{cancel}

\usepackage[normalem]{ulem}
\usepackage{float}
\usepackage{multirow}
\usepackage{amsfonts}
\usepackage{mathtools}
\usepackage{derivative}
\usepackage{braket}
\usepackage{physics}

\newcolumntype{P}[1]{>{\centering\arraybackslash}p{#1}}

\begin{document}

\title{
 Measurement-based quantum computation  with variable-range interacting systems}

\author{Debkanta Ghosh, Keshav Das Agarwal, Pritam Halder, Aditi Sen(De)}

\affiliation{Harish-Chandra Research Institute, A CI of Homi Bhabha National Institute,  Chhatnag Road, Jhunsi, Allahabad - 211019, India}

\begin{abstract}

We demonstrate that weighted graph states (WGS) generated via variable-range interacting Ising spin systems where the interaction strength decays with distance as a power law, characterized by the fall-off rate, can successfully implement single- and two-qubit gates with fidelity exceeding classical limits by performing suitable measurements. In the regime of truly long-range interactions (small fall-off rate), optimizing over local unitary operations, while retaining the local measurement scheme in the original measurement-based quantum computation (MBQC) set-up, enables the scheme to achieve nonclassical average  fidelities. Specifically, we identify a threshold fall-off rate of the interaction above which the fidelity of both universal single- and two-qubit gates consistently exceeds \(90\%\) accuracy. Moreover, we exhibit that the gate-implementation protocol remains robust under two realistic imperfections -- noise in the measurement process, modeled via unsharp measurements, and disorder in the interaction strengths. These findings confirm WGS produced through long-range systems as a resilient and effective resource for MBQC.  

\end{abstract}

\maketitle
\section{Introduction}
\label{sec:intro}

Breakthrough discoveries in quantum algorithms, which outperform their classical counterparts in solving various problems, have emphasized the need for quantum computers \cite{Feynman1986,nielsen_chuang_2010}. These algorithms are implemented through quantum circuits built from a universal set of quantum gates applied to an initial quantum state, followed by measurements \cite{divincenzo95,nielsen_chuang_2010}. Similar to classical computers, quantum error correction is crucial for mitigating the effects of noise and achieving fault-tolerant quantum computation. A key framework in this context is the stabilizer formalism,  \cite{Kitaev_1997, gottesman1997} which supports both quantum computation and communication by enabling error suppression below a critical threshold \cite{KITAEV20032}. Moreover, it also laid the groundwork for measurement-based quantum computation (MBQC) \cite{Rauss2001, rauss_pra2003, Hein2004, hein2006entanglement, casati2006quantum, browne2006oneway, briegel2009measurement}, an alternative model where computation is driven by quantum measurements rather than unitary evolution and it uses special classes of stabilizer states that are inherently robust against errors. 
A central example is the cluster state, a highly entangled state generated from a fully product state via nearest-neighbor (NN) Ising-type interactions, 
 and the universal quantum gates are realized via local quantum measurements and corresponding corrective unitaries.  Further studies have identified various states that can also serve as resources for MBQC \cite{grossnovel07, beyond1way07}, specifically showing that symmetry-protected topological states possess quantum computational power \cite{Nautrup_2015, Rauss_2017, Stephen_2017}.   

MBQC protocols are also studied with the generalization of cluster states, both in continuous variable \cite{Gu_2009, Ohliger_2012} and recently in discrete systems \cite{Kissinger2019, Descamps_2024, Wong2024}. A particular example of such states are weighted graph states (WGS) \cite{hartmann_briegel_2007}, which can be generated via long-range (LR) interactions in systems like trapped ions \cite{iontrap12, Islam_2011, Britton_2012} and cold atoms in optical lattices \cite{Lewenstein2007,Sowiński2012}, where spin interactions extend beyond NN. These states are shown to be highly entangled \cite{mahto_shaji_pra_2022, economou2023, Ghosh2024} and hence can potentially be useful for various quantum computation and information processing tasks such as ground state preparation \cite{Anders_2006,Anders_2007}, random number generation \cite{ran_circuit_by_wg08}, gate implementation \cite{Tame_2009} as well as interferometry \cite{peng12}. Further, LR models possess unique phenomena such as violation of the entanglement area law  \cite{cramer2010, koffel2012, scha2013, cadarso2013} and rapid propagation of correlation  \cite{Maghrebi2016,gong2017,ares2018,Ares_2019,lgcl_asd_pla2021}, and can encode information on the complete substrate as interactions \cite{Pons_2007}. While effects of such naturally present LR interactions can be suppressed \cite{liu2024}, WGS can offer insight into the LR properties of the evolving model \cite{dur_hartmann_prl_2005, Ghosh2024, ghosh2023}, and direct implementations of universal gates on WGS will enable quantum computation using non-stabilizer states. 

In this work, we examine the performance of  universal quantum gates  in the framework of the MBQC protocol when the system evolves under a long-range interacting Hamiltonian in which interactions decay with distance according to a power-law characterized by an exponent, \(\alpha\). Specifically, we adopt the same local projective measurements on the qubits as in the original MBQC \cite{browne2006oneway,Rauss2001}  after the evolution of the initial qubit state under the LR Hamiltonian. However, we optimize over all possible corrective unitaries applied to the qubits following measurement.
We exhibit that
even when the fall-off rate 
\(\alpha\) is moderate -- significantly lower than the NN limit (\(\alpha\rightarrow\infty\)) -- the resulting average gate fidelity can exceed the classical threshold \cite{Massar1995}, thereby guaranteeing the quantum advantage. 
We determine the minimum value of 
\(\alpha\) required for gate fidelity to reach a high accuracy level of  \(90\%\), both for the single- and two-qubit gate implementations. 
Interestingly, we find that for controlled-NOT($CNOT$) gates, the average fidelity consistently stays above the classical threshold across all values of the interaction strength. Moreover, within a specific range of \(\alpha <2\),  corresponding to genuinely long-range interactions, the fidelity notably surpasses \(75\%\), thereby ensuring benefits in the quantum domain through LR interactions.

During the implementation of measurement-based quantum computation, various types of errors and defects can arise at different stages of the protocol. One prominent source of error stems from imperfections in the measurement process, which may be influenced by environmental interactions with the measurement apparatus. A useful model to account for such noisy measurements is the concept of unsharp measurements \cite{Busch_2013,Silva15,Mal16}, where ideal projective measurements are admixed with white noise, thereby introducing a trade-off between the information gain from the measurement and the disturbance inflicted on the quantum state. Interestingly, unsharp measurements have been shown to be advantageous in various quantum information tasks, such as the discrimination of non-orthogonal states \cite{Peres1988, Huttner_1996, Derka_1998}, quantum control \cite{Gillett2010, Sayrin2011}, self-testing \cite{Tavakoli2020, Miklin2020}, and in the sharing of nonlocal correlations \cite{Silva15,Mal16,Hu2018, Gomez2018, Halder2022}, particularly in quantum networks \cite{Srivastava2021, Halder2021} and teleportation protocols \cite{Roy2021, Das2023}.
By analyzing the impact of unsharp measurements on WGS, we observe that the average gate fidelity decreases with increasing unsharpness for all values of $\alpha$ which can be attributed to residual entanglement in the post-measurement states. However, for low levels of unsharpness, the effect is minimal, and becomes more pronounced in systems with short-range interactions.

Additionally, imperfections may arise in the Hamiltonian used to generate entanglement, particularly when interaction strengths are site-dependent. Notably, such disordered systems can be engineered in a controlled manner in cold atomic platforms \cite{Ahuf2005,Lewenstein2007,Sanchez-Palencia2010,Jendrzejewski2012}. Despite potential drawbacks, the disorder has also been shown to offer robustness or advantages in some cases \cite{ahufinger06, Prabhu_2011, Shtanko_2020, ghosh_2020, konar_2022, konar_2022b, halder_2025}. Remarkably, our study finds that when site-dependent couplings are randomly drawn from a Gaussian distribution and averaged over multiple realizations, the quenched average gate fidelity remains largely unaffected, indicating strong robustness of the MBQC protocol against disorder.

The paper is structured in the following manner. In Sec. \ref{sec:sharp}, we introduce the protocol for the implementation of a single- and two-qubit gate with WGS through  measurements. The role of corrective unitaries on the average fidelity in the protocol with LR interacting Hamiltonian is discussed in Sec. \ref{sec:single}. The effects of unsharp measurement are discussed in Sec. \ref{sec:unsharp} while the investigations of disorder in the LR model are examined in Sec. \ref{sec:disorder}. Concluding remarks are included in Sec. \ref{sec:conclu}.

\begin{figure}
    \centering
    \includegraphics[width=0.95\linewidth]{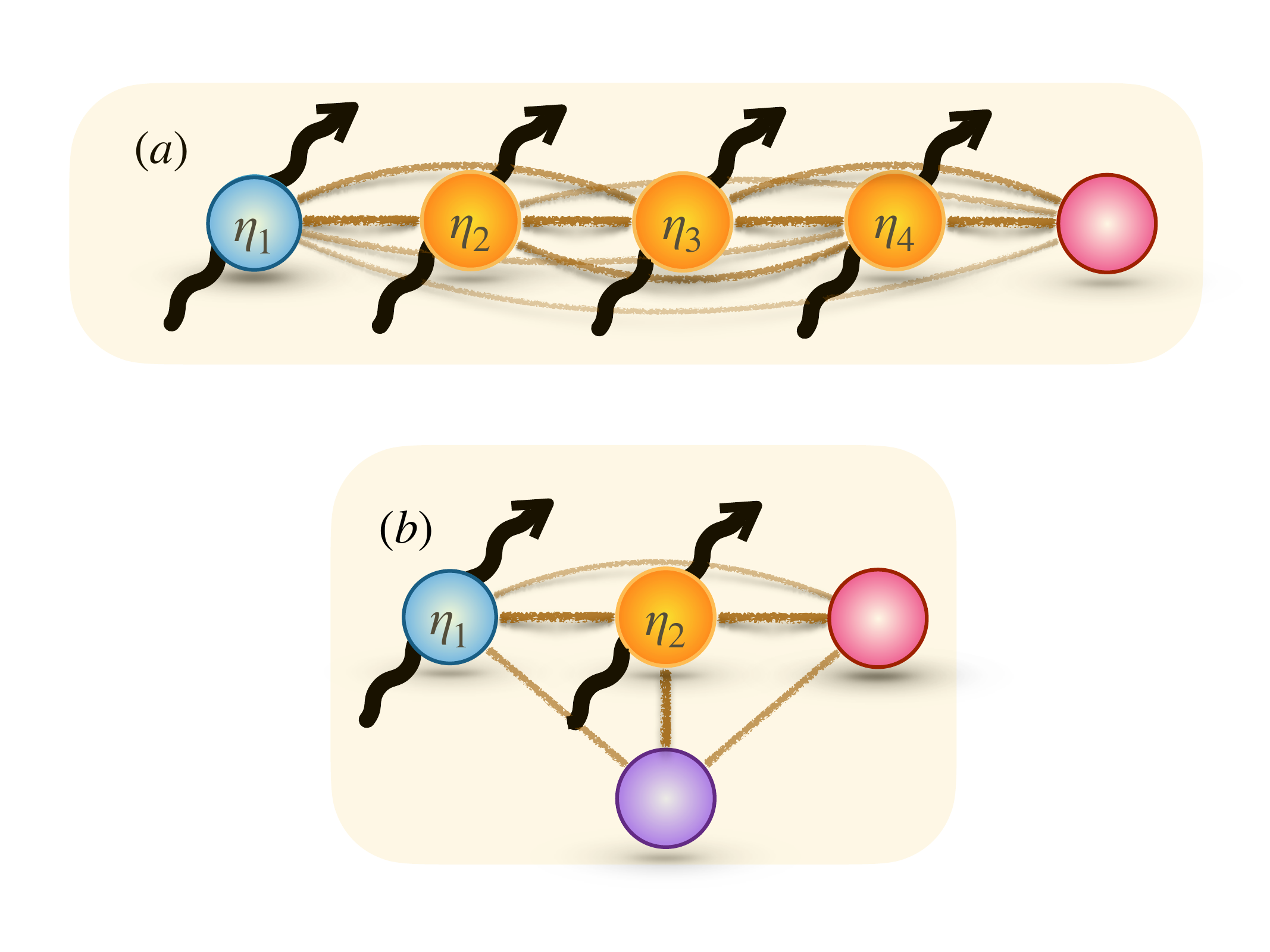}
    \caption{Schematic of a measurement-based quantum computation protocol on weighted graph state (WGS) created by using the LR interacting Hamiltonian. (a) A \(5\)-qubit WGS which can be part of a bigger cluster, \(\mathcal{W}(V,E)\), required for implementing a general single qubit gate. Blue and pink spheres represent the input  \((k\in V_I(\mathcal{G}))\) and output qubit \((k\in V_O(\mathcal{G}))\). The yellow ones signify the qubits, \(k\in V_M(\mathcal{G})\).   Measurements (both sharp and unsharp) are performed on the qubits, \(k\in V_m\) in the \(xy\)-plane, denoted by the black arrows, at an angle \(\eta_i)\) for \(i=1,2,3,4\). (b) A \(4\)-qubit WGS with a different geometry is used for implementing $CNOT$ (two-qubit) gate. The color code for qubits \(k\in V(\mathcal{G})\) is the same as in (a) except the violet one, which is the control qubit of the $CNOT$ gate that serves as both input and output. Note that only two measurements are required for implementing the $CNOT$ gate, different from the single-qubit gates.}
    \label{fig:schematic}
\end{figure}

\section{MBQC framework under a finite fall-off rate with unsharp measurements}
 \label{sec:sharp}
 
The cluster states (Appendix.~\ref{app:cluster}), belonging to an arbitrary lattice, $\mathcal C(V,E)$,  are the building blocks of measurement-based quantum computation \cite{briegel2009measurement,briegel01,Rauss2001}. Here, $V=\{i|i\in\mathcal C\}$ and $E=\{(i,j)|i,j\in \mathcal C,j\in \mathrm{nb}(i)\}$ are the set of vertices and nearest-neighbor edges ($\mathrm{nb}(i)$) of the cluster, respectively. In our case, since the evolving Hamiltonian is different from the original proposal \cite{Rauss2001}, the state produced is the weighted graph state, denoted as \(\mathcal{W}(V,E)\) instead of \(\mathcal{C}(V,E)\). To realize a gate, $U_\mathcal G$, the vertices $V(\mathcal G)$ of the WGS $\mathcal W(\mathcal G)$ are divided into three sets which include an input sector, $ V_I(\mathcal G)$, a body, $ V_M(\mathcal G)$, and an output part, $ V_O(\mathcal G)$, satisfying $ V_I(\mathcal G)\cup  V_M(\mathcal G)\cup  V_O(\mathcal G)= V(\mathcal G)$, $ V_I(\mathcal G)\cap V_M(\mathcal G)=\emptyset$, and $ V_M(\mathcal G)\cap  V_O(\mathcal G)=\emptyset$. Let us briefly discuss the  MBQC scheme to simulate $\mathcal G$ acting on the input state $\ket{\phi_{\text{in}}}$ with $\mathcal W(\mathcal G)$. We emphasize the steps in which it departs from the original proposal \cite{Rauss2001} referred to as the perfect MBQC (pMBQC) protocol.

{\it 1.  Initialization.}  Initialize the qubits as $\ket{\Phi_{\text{in}}}=\ket {\phi_{\text{in}}}_{V_I(\mathcal G)}\otimes \bigotimes_{i\in (V_M\cup V_O)\setminus V_I} \ket+_{i}$.

{\it 2. Evolving Hamiltonian and dynamical states. } The interacting Hamiltonian between the qubits, used to evolve the initial state, is given by \(H_\alpha=\sum_{k,l=1,k<l}^N H_{kl}\), where \(H_{kl}=g_{kl}\Big(\frac{1-\sigma_z^{(k)}}{2}\Big)\Big(\frac{1-\sigma_z^{(l)}}{2}\Big)\)  with \(\sigma_i \,(i=x,y,z)\) being the Pauli operators at sites \((k)\) and \((l)\),  and power-law interaction strength are represented as \(g_{kl}=\frac{J}{|k-l|^{\alpha}}\), with \(\alpha\) being the fall-off rate. Choosing $\alpha\to\infty$ makes $H_{\alpha\to\infty}\to H_{NN}=\sum_{k,l=1,k<l,|k-l|=1}^N \Big(\frac{1-\sigma_z^{(k)}}{2}\Big)\Big(\frac{1-\sigma_z^{(l)}}{2}\Big)$ as the nearest-neighbor interacting Hamiltoninan, leading to the cluster states, \(\mathcal{C}(V,E)\) at time, \(t= n\pi\) \( \forall \, n \in Z^{+}\). On the other hand, the evolving Hamiltonian \(H_\alpha\) creates a class of genuinely multipartite entangled states, \(\ket{\Phi^e(\alpha)}\), at time \(t= \pi\), which changes its characteristics depending on the fall-off rate \(\alpha\) \cite{ghosh2023}. We will illustrate that the presence of variable-range interactions has nontrivial impacts on MBQC. Moreover, imperfections or disorder naturally appear in quantum many-body systems or can be artificially introduced in systems with ultracold atomic gases in optical lattices. It can be modelled by considering site-dependent coupling \(\{J_i\}\), with \(i\) denoting the sites,  chosen from a random Gaussian distribution with a specific mean, \(\bar{J}\), and standard deviation \(\sigma\). The influence of the disorder along with the variation of \(\alpha\) on MBQC will be scrutinized in detail.

{\it 3. Local measurements for creating outputs.} First, we define the set of qubits $V_m=(V_I(\mathcal G)\cup  V_M(\mathcal G))\setminus V_O$ entitled to  measurements. The local projective measurements (PV) on the qubits $k\in V_m$ of the evolved state $\ket{\Phi^e(\alpha)}$ are performed in the basis \(\ket{\eta_{s_k}}=\frac{\ket{0}_k+e^{\iota(s_k\pi+\eta_k)}\ket{1}_k}{\sqrt{2}}\), (where $\iota = \sqrt{-1}$). Here \(s_k\in\{0,1\}\) are the measurement outcome for the \(k\)th qubit and \(\eta_k\in[0,2\pi]\) is the  angle between the Bloch vector in the \((x,y)\)-plane and the positive \(x-\)axis  of the Bloch sphere. Therefore, the total measurement operator is characterized by $ \mathbf M_{\mathbf s,\mathcal G}(\{\eta_{s_k}\}_{k\notin V_O})=\bigotimes_{k\in V_m}\ketbra{\eta_{s_k}}$, where $\mathbf s=\{s_k\}_{k\in V_m}$. The local projective measurements performed on the qubits by a measurement apparatus are typically disturbed by noise \cite{Busch_2013,Silva15,Mal16,Shenoy19}. In particular, when a measurement is performed indirectly through an auxiliary system that interacts with the qubit of interest, the duration of the interaction presents a trade-off between the invasiveness of the measurement and the disturbance it causes.  The "unsharpness" on the projective measurement on the $k$th qubit can be effectively modeled using a convex combination of a projective measurement and the maximally mixed state (white noise), given by
\begin{equation}
    P_{s_k}^\lambda = \lambda \ketbra{\eta_{s_k}}{\eta_{s_k}} + (1 - \lambda)\frac{{I}}{2},
\end{equation}
where \( 0 \leq \lambda \leq 1 \) is referred to as the \emph{unsharpness parameter} with $(1-\lambda)$ quantifying the degree of noise in the measurement. The set \( \{P_{s_k}^\lambda\}_{s_k=0}^1 \) forms a positive operator-valued measure (POVM) satisfying the completeness relation \( \sum_{s_k=0}^1 P_{s_k}^\lambda = {I} \forall \lambda\). Given a state $\varrho$ as an input, the unnormalized output state corresponding to  $P_{s_k}^\lambda$ is given by $\sqrt{P_{s_k}^\lambda}\varrho\sqrt{P_{s_k}^\lambda}$. Note that perfect measurements are required for information processing in MBQC, so to disentangle the state perfectly, while these unsharp measurements are unable to disentangle the qubits from the output sector, thereby affecting the performance of the MBQC scheme. Assuming equal noise strength up to $n$th qubit local measurements \((n=1,2,\dots,|V_m|)\) along which information flow may occur, the total measurement operator $\mathbf{M}_{\mathbf s,\mathcal G}(\{\eta_{s_k}\}_{k\notin V_O})$ can now be updated as
\begin{eqnarray}
   \nonumber &&\mathbf{M}_{\mathbf s,\mathcal G}^{\lambda,n}(\{\eta_{s_k}\}_{k\notin V_O})\\&&=\bigotimes_{\substack{k_1=1\\k_1\in V_m}}^n\sqrt{P_{s_{k_1}}^\lambda}\otimes\mathbf{M}_{\mathbf s,\mathcal G}\left(\left\{\eta_{s_{k_2}}\right\}_{\substack{k_2=n+1\\k_2\in V_m}}^{|V_m|}\right),
\end{eqnarray}
where $n\leq|V_m|$. Under these circumstances, the average output state on $V_O$ can be written as $\Lambda_{\alpha}^{\lambda,n}(\ketbra{\phi_\text{in}})$, where $\Lambda_{\alpha}^{\lambda,n}$ is a completely positive trace-preserving map (CPTP) defined by
\begin{eqnarray}
    \Lambda_{\alpha}^{\lambda,n}(\ketbra{\phi_\text{in}}) = \text{Tr}_{V_m}\big(\sum_\mathbf{s}\mathbf{M}_{\mathbf{s},\mathcal G}^{\lambda,n}\ketbra{\Phi^e_\alpha}\mathbf{M}_{\mathbf{s},\mathcal G}^{\lambda,n}\big).
\end{eqnarray}
    
{\it 4. Optimal local corrective unitaries for gate implementation. } In the case of  PV measurements performed on \(V_m\), the output state on $\mathcal W(\mathcal G)$ corresponding to the outcome $\mathbf s$ is given by \( \ket{\Phi_{\text{out}}} = \left(\bigotimes_{k\in V_m}\ket{\eta_{s_k}}\right)\otimes \ket{\phi_\text{out}}_{V_O}\).  In the outcome sector, the state has the form $  \ket{\phi_\text{out}}_{V_O} = U_{c,\mathcal G}^\mathbf{s}U_{\mathcal G}\ket{\phi_\text{in}}$ for   $\alpha\to\infty$. Here, $U_{c,\mathcal G}^\mathbf{s}=\otimes_{k\in V_m}(\sigma_x^k)^{x_k}(\sigma_z^k)^{z_k}$ is the local corrective unitary where the explicit form of $x_k,z_k\in\{0,1\}$ are calculated with respect to the output $\mathbf s$, depending on $U_\mathcal{G}$. Note that $\sigma_i^k$ denotes the application of the Pauli spin operator $\sigma_i$ on the $k$th qubit. Since each $U_{c,\mathcal G}^\mathbf{s}$ is a tensor product of Pauli matrices, we get the desired output state, $U_{\mathcal G}\ket{\phi_\text{in}}$ by applying $U_{c,\mathcal G}^\mathbf{s}$ on $\ket{\phi_\text{out}}$. Further, we notice that the local unitary operators which are optimal for the cluster state may not be so for the WGS, and hence, depending on \(\alpha\), optimization over local unitaries has to be performed. Moreover, when measurements are noisy, the resulting output state depends both on $\alpha$ and $\lambda$, and hence the corrective unitaries  $U_{c,\mathcal G}^{\alpha,\mathbf{s}}$ have to be chosen suitably depending on the outcome. Therefore, we find 
\begin{eqnarray}
    \nonumber \rho_\text{out}&=&\Lambda_{\alpha,\mathcal G}^{\lambda,n}(\ketbra{\phi_\text{in}}) \\&=& \text{Tr}_{V_m}\big(\sum_\mathbf{s}U_{c,\mathcal G}^{\alpha,\mathbf{s}}\mathbf{M}_{\mathbf{s},\mathcal G}^{\lambda,n}\ketbra{\Phi^e_\alpha}\mathbf{M}_{\mathbf{s},\mathcal G}^{\lambda,n} U_{c,\mathcal G}^{\alpha,\mathbf{s}}\big),
\end{eqnarray}
where
\begin{equation}
   U_{c,\mathcal G}^{\alpha,  \mathbf{s}} = (\sigma_x)^{c_0+\sum_{k\notin V_O}c_ks_k}(\sigma_z)^{d_0+\sum_{k\notin V_O}d_ks_k}, 
   \label{eq:Uc_general}
\end{equation}
with \( c_k, d_k \in \{0,1\} \)  \(\forall k \) being functions of $\alpha$ and $\lambda$ in general.

 \textit{Performance quantifier.} The quality of the MBQC protocol is quantified by the maximum achievable fidelity between $\rho_\text{out}$ and $U_\mathcal{G}\ket{\phi_\text{in}}$. Mathematically, it takes the form as 
\begin{eqnarray}
    \nonumber\mathcal{F}_{\mathcal{G}}(\Lambda_{\alpha,\mathcal G}^{\lambda,n},U_{\mathcal{G}})=\max_{U_{c,{\mathcal{G}}}^{\alpha,\mathbf{s}}}\bra{\phi_\text{in}}{U_{\mathcal{G}}^{\dagger}\Lambda_{\alpha,\mathcal G}^{\lambda,n}(\ketbra{\phi_\text{in}})U_{\mathcal{G}}\ket{\phi_\text{in}}},\\
     \label{eqn:fidelity}
\end{eqnarray}
where both the fall-off rate and unsharp measurement are incorporated. To remove the input state-dependency, the average gate fidelity,
\begin{eqnarray}
    \bar{\mathcal{F}}_{\mathcal{G}}(\Lambda_{\alpha,\mathcal G}^{\lambda,n},U_{\mathcal{G}})=\frac{\int\mathcal{F}_{\mathcal{G}}(\ket{\phi_\text{in}})\,d\ket{\phi_\text{in}}}{\int d\ket{\phi_\text{in}}}.
    \label{eqn:avg_fidelity_f}
\end{eqnarray}
is computed. Moreover, Eq.~\eqref{eqn:avg_fidelity_f} reduces to \cite{NIELSEN2002avgfid}
\begin{eqnarray}
   \nonumber \bar{\mathcal{F}}_{\mathcal{G}}(\Lambda_{\alpha,\mathcal G}^{\lambda,n},U_{\mathcal{G}})=\frac{d^2+\sum_{i=0}^{d^2-1 }\Tr(U_{\mathcal{G}}U_i^{\dagger}U_{\mathcal{G}}^{\dagger}\Lambda_{\alpha,\mathcal G}^{\lambda,n}(U_i))}{d^2(d+1)},\\
    \label{eq:fid_gen}
\end{eqnarray}
where \(d\) is the dimension of the input states and $\big\{U_i/\sqrt d\big\}_{i=0}^{d^2-1}$ forms an orthonormal operator basis in $\mathbb C^d$ satisfying Hilbert-Schmidt inner product, i.e., $\Tr(U_i^\dagger U_j)=\delta_{ij}d$. We are now ready to demonstrate the effect of fall-off rate, $0\leq\alpha\leq\infty$, and $\lambda$ on the performance of MBQC. For brevity, when $\lambda=1$, we denote the average gate fidelity in Eq.~\eqref{eq:fid_gen} simply as $\bar{\mathcal{F_G}}$.

\section{Effects of Fall-off rates  on fidelity for implementing  gates}
\label{sec:single}

The involvement of a long-range interacting Hamiltonian in evolution changes the features of the time evolved state qualitatively, and hence it is obvious that the performance of computation gets affected. To exhibit its impact, we explore the single- and two-qubit gates required for universal quantum computation \cite{nielsen_chuang_2010}.

\subsection{Functioning of single-qubit gates with LR interactions}

The general single-qubit rotations, $U_{\mathcal G}\in SU(2)$ can be realized on a chain of five-qubits with $V_I=\{1\}$, $V_M=\{2,3,4\}$ and $V_O=\{5\}$ as shown in Fig.~\ref{fig:schematic}. It is possible to decompose any rotation \(U_{\mathcal G}\) as
\begin{eqnarray}
 U_{\mathcal G}(\theta_1,\theta_2,\theta_3,\theta_4)=e^{-\iota\theta_1}R_x(\theta_2)R_z(\theta_3)R_x(\theta_4),
\end{eqnarray}
where $R_j(\theta)=e^{-\iota\sigma_i \theta/2}$ $\forall i=x,y,z$. To simulate $U_{\mathcal G}(\theta_1,\theta_2,\theta_3,\theta_4)$ in the MBQC process, the measurement angles take the form \(\eta_1=0,\,\eta_2=\theta_2(-1)^{s_1+1},\,\eta_3=-\theta_3(-1)^{s_2},\,\eta_4=-\theta_4(-1)^{s_1+s_3}\). We now analyze the impact of a finite fall-off rate on the fidelity of a set of universal single-qubit gates, specifically the Hadamard gate ($H$), the $\pi/2$-phase gate ($U_{R_z(\pi/2)}$), and the $T$-gate. Note that, in the qubit scenario, by choosing Pauli matrices $\{I,\sigma_x,\sigma_y,\sigma_z\}$ as the operator basis, Eq.~\eqref{eq:fid_gen} transforms to 
\begin{eqnarray}
    \bar{\mathcal{F}}_{\mathcal{G}}=\frac{1}{2}+\frac{1}{12}\sum_{i=1,2,3}\Tr(U_{\mathcal{G}}\sigma_i U_{\mathcal{G}}^{\dagger}\Lambda_{\alpha,\mathcal G}^{\lambda,n}(\sigma_i)).
    \label{eq:fid_qubit}
\end{eqnarray}
The optimal corrective unitary in the limit $\alpha\to\infty$ and $\lambda=1$ match those used in pMBQC \cite{rauss_pra2003}. However, we will show that for smaller values of \( \alpha \), corrective unitaries become $\alpha$-dependent.

\begin{figure}
    \centering
    \includegraphics[width=\linewidth]{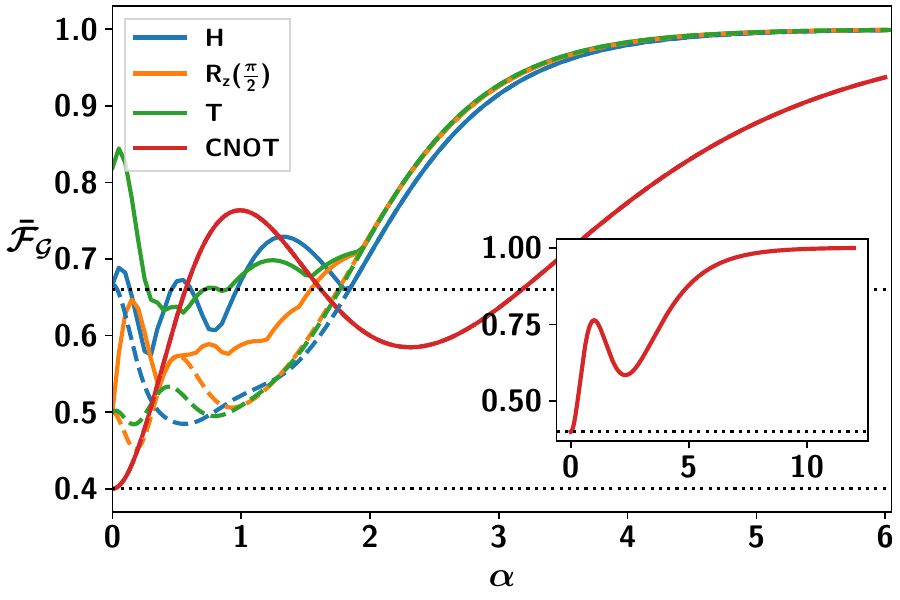}
    \caption{(Color online.) The average fidelity, \(\bar{\mathcal{F}}_{\mathcal{G}}\) (vertical axis) of different gates (single-qubit and two-qubit) against the fall-off rate \(\alpha\) (horizontal axis). Different solid lines represent average fidelities for different gates after maximizing over corrective unitaries $U_{c,\mathcal{G}}^{\alpha,\mathbf{s}}$ for a particular \(\alpha\). The dashed lines are for  \(\bar{\mathcal{F}}^r_{\mathcal{G}}\) calculated using the set of corrective unitaries $U_{c,\mathcal{G}}^{\infty,\mathbf{s}}$ as in the pMBQC protocol.  Comparing solid and dashed lines, it is clear that the maximization of corrective unitaries is essential for small \(\alpha\) values.   The black dotted lines mark the classical fidelity limits \(\mathcal{F}_c=2/3\) and \(1/2\) for a single- and two-qubit gates, respectively. Note that \(\bar{\mathcal{F}}_{\mathcal{G}}\) starts from a finite value and reach to unity with an accuracy \(\mathcal{O}(10^{-3})\) after \(\alpha\gtrsim 4.5\) for single-qubit gate, where as \(\alpha\gtrsim 8.66\) for a  $CNOT$ gate. The inset highlights the trend for the $CNOT$ gate, up to \(\alpha=12\). Both axes are dimensionless.} 
    \label{fig:sharp}
\end{figure}

{\bf Hadamard gate.} First, note that Hadamard gate  can be decomposed as  \(U_{H}(\frac{\pi}{2},\frac{\pi}{2},\frac{\pi}{2},\frac{\pi}{2})= e^{-\iota\frac{\pi}{2}}e^{-\iota\frac{\pi}{4}\sigma_x}e^{-\iota\frac{\pi}{4}\sigma_z}e^{-\iota\frac{\pi}{4}\sigma_x}\). The measurement operator to simulate $H$ reads as \(\mathbf M_{\mathbf s,H}(0,-\frac{\pi}{2}(-1)^{s_1+1},-\frac{\pi}{2}(-1)^{s_2},-\frac{\pi}{2}(-1)^{s_1+s_3})\). Here we choose the same measurements as in pMBQC. Although we numerically verify that other local measurements do not help to increase the average fidelity. The fidelity $\bar{\mathcal F}_H$ can be obtained from Eq.~\eqref{eq:fid_qubit} by optimizing over corrective unitaries (see Fig.~\ref{fig:sharp}). In the regime $\alpha\gtrsim1.82$, the fidelity is a smooth function of $\alpha$, always surpassing the classical fidelity, $F_c=2/3$ \cite{Massar1995,tele_fid99}, and saturate to unity (up to $\mathcal O(10^{-3})$) for $\alpha\gtrsim4.5$ which we denote as $\alpha_s$ (see Table.~\ref{tab:sharp}). We can safely assume that $\alpha\gtrsim 5$ mimics the NN model, thereby attaining the unit fidelity.  Interestingly, however, we find that optimal corrective unitaries $U_{c,H}^{\alpha\gtrsim1.82,\mathbf{s}}=\sigma_x^{s_2+s_4}\sigma_z^{s_1+s_3+1}$  are same as those for pMBQC, i.e., $U_{c,H}^{\infty,\mathbf{s}}$, for $\alpha\gtrsim1.82$. Although when $\alpha<1.82$, unitaries which optimize the average fidelity depend on $\alpha$ and are different for each $\alpha$. In this regime, $\bar{\mathcal F}_H$ is highly fluctuating, and the same measurement strategy with optimal unitaries does not beat the classical fidelity. If we apply the same corrective unitaries used for the pMBQC protocol,   the average fidelity for all \(\alpha\) reads as 
 \begin{eqnarray}
    \nonumber \bar{\mathcal F}_H^r &=& \frac{1}{48} \Big( 28 + 2\cos a_2 + 4\cos b_3\\ \nonumber
    &+& 2\cos2a_2\cdot[\cos a_4+\cos b_4]\\ \nonumber
    &+&\cos a_2\cdot[3\cos a_4+2\cos b_3+4\cos b_4\\ 
    &+& \cos(b_3+b_4)]
    \Big)=\bar{\mathcal F}_H (\alpha\gtrsim 1.82),
    \label{eqn:fid_H}
    \end{eqnarray}
where $a_j=j^{-\alpha}\pi$ for $j=2,3,4$, $b_3=a_2+a_3$ and $b_4=b_3+a_4$. Notice first that $\bar{\mathcal F}_H^r$ matches $\bar{\mathcal F}_H$ beyond $\alpha\approx1.82$, implying that the unitaries for the pMBQC protocol are suitable for realizing the Hadamard gate even for the LR interacting model. Secondly, the superscript $``r"$ represents the restrictive scenario for \(\alpha <1.82\) since it emphasizes the requirement to optimize over corrective unitaries in this truly LR domain. Note that although $\bar{\mathcal F}_H^r$ never goes beyond classical fidelity for $\alpha<1.82$, optimizing over corrective unitary can achieve maximum average gate fidelity $\bar{\mathcal F}_H=0.72$ at $\alpha=1.32$ in this region, which again highlights the $\alpha$-dependence of the fidelity when evolving Hamiltonian depends on $\alpha$ (see Table.~\ref{tab:sharp} where we note the maximum average gate fidelity, $\bar{\mathcal F}_H^{\max}$, achieved in the presence of $\alpha$ when $\alpha\in[0,2)$, denoted as $\alpha_{\max}$). We also indicate the minimum $\alpha$-value, $\alpha_{\min}^{th}$, for which the average gate fidelity reaches the threshold of $90\%$ accuracy, referred to as $\bar{\mathcal F}_H^{th}$ (see Table.~\ref{tab:sharp}).   We choose this precision since the fidelity of single-qubit gates in currently available architectures \cite{Wright2019}  can achieve and exceed that limit.  Further, we observe that such a situation occurs when the fidelity function becomes smooth and $\alpha \in (2,3)$.

    

    \begin{table}[]
        \centering
        \begin{tabular}{|c|cc|c|c|}
        \hline
           \multirow{2}{*}{Gate (\(\mathcal{G}\))}  & \multicolumn{2}{|c|}{\(0\leq\alpha\leq2\)} & &\\
           & \(\bar{\mathcal{F}}_{\mathcal{G}}^{\text{max}}\)&\(\alpha_{\text{max}}\)&\(\alpha_s\) &\(\alpha^{th}_{\text{min}}\)\\ \hline
            $H$ &\(0.72\)&\(1.32\)&4.5&2.89\\\hline
            \(\frac{\pi}{2}\)-phase&\(0.73\)&\(1.99\)&4.5&2.8 \\\hline
             $T$ & \(0.84\)&\(0.05\)&4.5&2.78\\\hline
             $CNOT$ & \(0.76\)&\(1.00\)&8.66&5.31\\\hline
            
        \end{tabular}
        \caption{The first column indicates the gates, \(\mathcal{G} \in \{H, \pi/2-\text{phase}, T,CNOT\}\) considered in this work. The other columns represent the maximum achievable fidelity \(\bar{\mathcal{F}}_{\mathcal{G}}^{\max}\) and its corresponding fall-off rate \(\alpha_{\max}\) when \(0 \leq \alpha  < 2\), the \(\alpha\)-values where the fidelity saturates and the threshold value of \(\alpha\) above which \(\bar{\mathcal{F}}_{\mathcal{G}} \geq 0.9\) for different gates.}
        \label{tab:sharp}
    \end{table}

{\bf $T$ gate and \(\frac{\pi}{2}\)-phase gate:}
A \(Z\)-rotation at  specific angles \(\frac{\pi}{2}\) and \(\frac{\pi}{4}\) are known as \(\frac{\pi}{2}\)-phase gate, $U_{R_z(\frac{\pi}{2})}(0,0,\frac{\pi}{2},0)=e^{-\iota\frac{\pi}{4}\sigma_z}$  and $T$ gate, $U_T(0,0,\frac{\pi}{4},0)=e^{-\iota\frac{\pi}{8}\sigma_z}$, respectively. Below $\alpha\approx1.77$, the average gate fidelity $\bar{\mathcal F}_{T(R_z(\pi/2))}^r\leq F_c$ after applying the same local measurements and corrective unitaries on $V_m$ and $V_O$ respectively as in pMBQC \cite{rauss_pra2003}. However, the maximization over corrective unitaries lead to \( \bar{\mathcal{F}}_T \) and \( \bar{\mathcal{F}}_{R_z(\pi/2)} \) attaining their respective maximum values of $0.84$ and $0.73$ at \( \alpha_{\max} = 0.05 \) and \( \alpha_{\max} = 1.99 \), respectively. For $\alpha\gtrsim1.93$, both the fidelity $\bar{\mathcal F}_{R_z(\pi/2)}$ and $ \bar{\mathcal F}_T$ have optimal corrective unitary $U_{c,{R_z(\frac{\pi}{2})}}^{\alpha\gtrsim1.93,\mathbf{s}}= U_{c,T}^{\alpha\gtrsim1.93,\mathbf{s}}=\sigma_x^{s_2+s_4}\sigma_z^{s_1+s_3}=U_{c,{R_z(\frac{\pi}{2})}}^{\infty,\mathbf{s}}=U_{c,T}^{\infty,\mathbf{s}}$, again coinciding with  the NN case. The fidelity expression for \(\frac{\pi}{2}\)-phase gate in this restrictive case can be computed as
\begin{eqnarray}
\nonumber   \Bar{\mathcal{F}}^r_{R_z(\frac{\pi}{2})}&=& \frac{1}{48} \Big( 
27 + \cos 2a_2 + 3\cos a_2 +3\cos a_3 \\ \nonumber
&+&2 \cos b_3 + \cos(b_3+a_2) + \cos(b_3+a_3) \\ \nonumber
&+&2\cos a_2 \cdot[\cos(a_2+a_4) + \cos(a_3+a_4) \\
&+&\cos a_3+\cos b_4 + \cos(b_3+b_4)]
\Big),
\label{eqn:fid_phase}
\end{eqnarray}
which become  $\Bar{\mathcal{F}}_{R_z(\frac{\pi}{2})}$ for $\alpha\gtrsim 1.93$. A similar argument can be made for $\Bar{\mathcal{F}}^r_{T}$ whose detailed expression is given in Eq.~\eqref{eqn:fid_T} of Appendix.~\ref{app:tgate}. We also find that both $\Bar{\mathcal{F}}^{th}_{R_z(\frac{\pi}{2})}$ and $\Bar{\mathcal{F}}^{th}_T$ achieve $0.9$, when $\alpha_{\min}^{th}\approx2.8$ and $\approx2.78$ respectively.


The entire analysis of the single-qubit gate implementations provides a few important insights -- $(1)$ the unit fidelity can be achieved for a finite value of $\alpha(\gtrsim4.5)$; $(2)$ optimizing over the corrective unitaries is essential for small $\alpha$, especially for $\alpha<2$ to beat the classical limit; $(3)$ the same measurement framework along with optimal local unitaries can provide quantum benefits even when LR interacting Hamiltonian is involved in the evolution.


\subsection{Implementation of $CNOT$ gate}
Two-qubit controlled-NOT gate, along with the previously discussed single-qubit gates, can perform universal quantum computation. To simulate $CNOT$,  we consider four qubits in a specific geometry, as shown in  Fig.~\ref{fig:schematic}. Here, we have $V_I=\{1,4\}$, $V_M=\{2\}$ and $V_O=\{3,4\}$ along with measurement operator \(\mathbf M_{\mathbf s, CNOT}(0,0)\). Utilizing the Heisenberg-Weyl operator basis, \(\{U_{ij}=\omega^{-\frac{ij}{2}Z^iX^j}\}_{i,j=0}^3\) with \(\omega=\exp(\frac{2\iota\pi}{4})\), \(Z=\sum_{i=0}^3\omega^i\ketbra{i}{i}\) and \(X=\sum_{i=0}^3\ketbra{i-1}{i}\) (modulo \(3\)) in $d=4$ we have $\bar{\mathcal{F}}_{CNOT}=\frac{1}{5}+\frac{1}{80}\sum_{i,j=0}^3\Tr(U_{\mathcal{G}}U_{ij} ^{\dagger}U_{\mathcal{G}}^{\dagger}\Lambda_{\alpha,\mathcal G}^{\lambda,n}(U_{ij}))$. 

In this situation, some unique features emerge that are distinct from the single-qubit gates observed with a variable-range interacting Hamiltonian. Firstly, $\bar{\mathcal{F}}_{CNOT}>F_c= 0.4$, for all values of $\alpha$ although $\bar{\mathcal{F}}_{CNOT}$ behaves nonmonotonically with the increase of $\alpha$. Secondly, $\bar{\mathcal{F}}_{CNOT}$ achieves its local maximum $0.76$ at $\alpha_{\max}\approx0.99<2$. Thirdly, it monotonically increases when $\alpha\gtrsim2.5$ as shown in  Fig.~\ref{fig:sharp}. Further, it attains to unit fidelity, thereby reproducing the results with the NN model $(\alpha\to\infty)$ for $\alpha\gtrsim8.66$, which is much higher than the ones obtained for a single-qubit case. Finally, in the case of the $CNOT$ gate, we notice that, $\bar{\mathcal{F}}_{CNOT}=\bar{\mathcal{F}}_{CNOT}^r$ where the optimal corrective unitary  \(U_{c,CNOT}^{\alpha,\mathbf{s}}={\sigma_z^{(3)}}^{s_1}{\sigma_x^{(3)}}^{s_2}{\sigma_z^{(4)}}^{s_1}\) turns out to be independent of $\alpha$.


\section{Imperfect measurement in MBQC with long-range interaction}
\label{sec:unsharp}

\begin{figure}
    \centering
    \includegraphics[width=\linewidth]{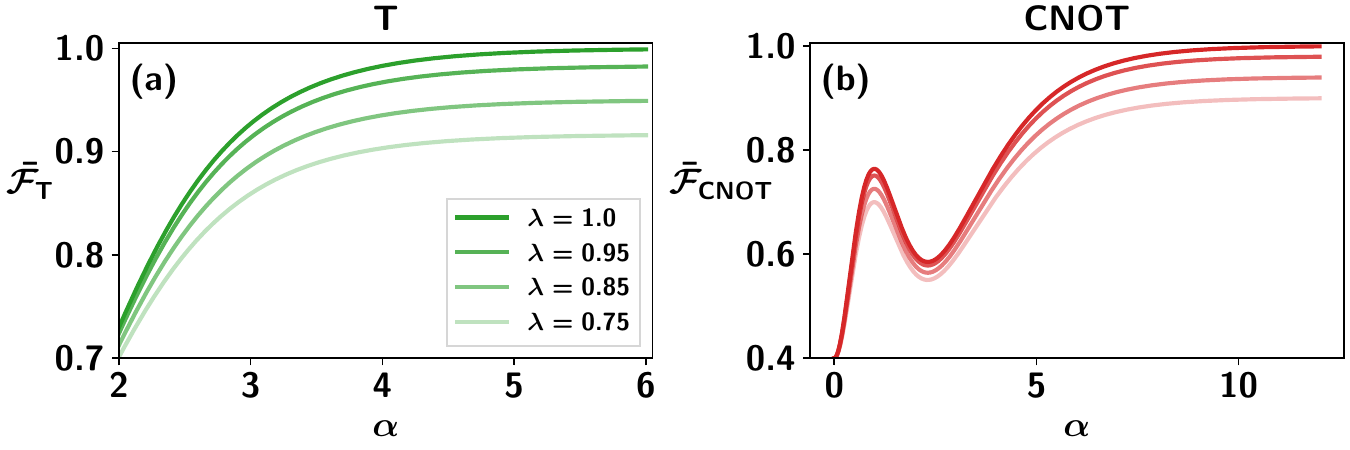}
    \caption{(Color online.) The average fidelity, \(\bar{\mathcal{F}}_{\mathcal{G}}\) (abscissa) against fall-off rate $\alpha$ (ordinate) for different unsharp parameter \(\lambda\). (a)  Fidlities for \(T\) and (b) $CNOT$ gates. Dark to light shades in (a) and (b) represent the decrease of the unsharp parameter \(\lambda\). Effects of \(\lambda\) are negligible around  \(\alpha < 2\) although difference emerges for different \(\lambda\) after \(\bar{\mathcal{F}}_{\mathcal{G}}\) starts increasing before saturation to a finite value \(\alpha\gtrsim 4.5\) for \(T\) and \(\alpha\gtrsim8.66\) for $CNOT$ gates. Both axes are dimensionless.} 
    \label{fig:unsharp_one_qubit}
\end{figure}

Let us now study the influence of unsharp measurement on MBQC along with long-range interaction. Of course, when $\lambda=1$ for all measurements and $\alpha\to\infty$, it is a pMBQC protocol. On the other hand, if the unsharp measurement is performed on the first qubit while the sharp measurements are applied on the rest of the qubits, surely, we have $\bar{\mathcal{F}}_{\mathcal{G}}\geq \bar{\mathcal{F}}_{\mathcal{G}}(\Lambda^{\lambda< 1,n}_{\alpha,\mathcal G})$. Let us assume at this point that although measurements are noisy, corrective local unitary found for each $\alpha$ value in the case of PV measurements will be used to compute the average gate fidelity. This is a plausible assumption as the set-up for the entire scheme is arranged before executing the process in which noise or defects enter. Moreover, we initially fix that only the first qubit is affected by the noisy measurement and hence, following Eq.~\eqref{eq:fid_gen}, we compute $\bar{\mathcal{F}}_{\mathcal{G}}(\Lambda_{\alpha,\mathcal G}^{\lambda,1},U_{\mathcal{G}})$. For both single- and two-qubit gates, we first observe that $\bar{\mathcal{F}}_{\mathcal{G}}(\Lambda_{\alpha,\mathcal G}^{\lambda,1},U_{\mathcal{G}})$ monotonically decreases with the decrease of $\lambda$ with the variation of $\alpha$ (see Fig.~\ref{fig:unsharp_one_qubit}).  As noted before and as expected, the average fidelity is almost close to unity when $\lambda=1$, especially for $\alpha\gtrsim 4.5$ (for single qubit gates) and $\alpha\gtrsim 8.66$ (for two-qubit gates). 

Let us now analyze the situation when \(\lambda <1\), which reveals how unsharpness parameters in measurements can affect the process. To this aim, we introduce the quantity,  the robustness of fidelity (RF), defined as  
\begin{eqnarray}
&&\nonumber\Delta^n\bar{\mathcal{F}}_{\mathcal{G}}^{\lambda}\\\nonumber&&= \frac{\max(\bar{\mathcal{F}}_{\mathcal{G}}(\Lambda_{\alpha,\mathcal G},U_{\mathcal{G}}),F_c)-\max(\bar{\mathcal{F}}_{\mathcal{G}}(\Lambda_{\alpha,\mathcal G}^{\lambda,n},U_{\mathcal{G}}),F_c)}{\max(\bar{\mathcal{F}}_{\mathcal{G}}(\Lambda_{\alpha,\mathcal G},U_{\mathcal{G}}),F_c)} \times 100\%,\\
\end{eqnarray}
where the first term  in the numerator and denominator is for the PV measurements, while the second term in the numerator characterizes the involvement of unsharp measurements upto $n$ qubits (\(n=1,2, \ldots ,|V_m|\). This quantity can address the question --``Is the decrease of fidelity with the unsharp parameters, non-monotonic with the variation of $\alpha$?" A behavior of $\Delta^n\bar{\mathcal{F}}_{\mathcal{G}}^{\lambda}$ with $\alpha$ may reveal this non-monotonicity, highlighting the varying noise resilience against $\lambda$ with LR interactions. Indeed, we exhibit that unsharpness of the measurement and the LR interactions nonlinearly affects the performance of the gate fidelity.

Let us first discuss the result for \(n=1\). Our numerical simulations reveal that the $x\%$ unsharpness in the measurements results in approximately $\frac x2\%$ decrease in the average fidelity in the case of single-qubit gates. On the other hand, when $\alpha$ is small, $\alpha\lesssim3$, the impact of $\lambda$ on $\bar{\mathcal{F}}_{\mathcal{G}}$ is not that substantial, eg., when $\alpha=2.5$, $\bar{\mathcal{F}}_{T}=0.84,0.82$ and $0.80$ for $\lambda=0.95, 0.85$ and $0.75$ respectively which is approximately  $\frac x4\%$ decrease of the original value with \(\lambda =1\). A similar observation can also be drawn for the $H$ and $\pi/2$-phase gate.  In the case of the $CNOT$ gate, we choose two $\alpha$ values - (i) the $\alpha$-value where the local minimum is achieved, $\bar{\mathcal{F}}_{{CNOT}}=0.74$ with $\lambda=0.95$ (the decrease of \(2\%\)) and (ii) at $\alpha_s, \bar{\mathcal{F}}_{{CNOT}}=0.97$ with  $\lambda=0.95$ which implies \(3\%\) of  decrease from the sharp measurement case.

\begin{figure}
    \centering
    \includegraphics[width=\linewidth]{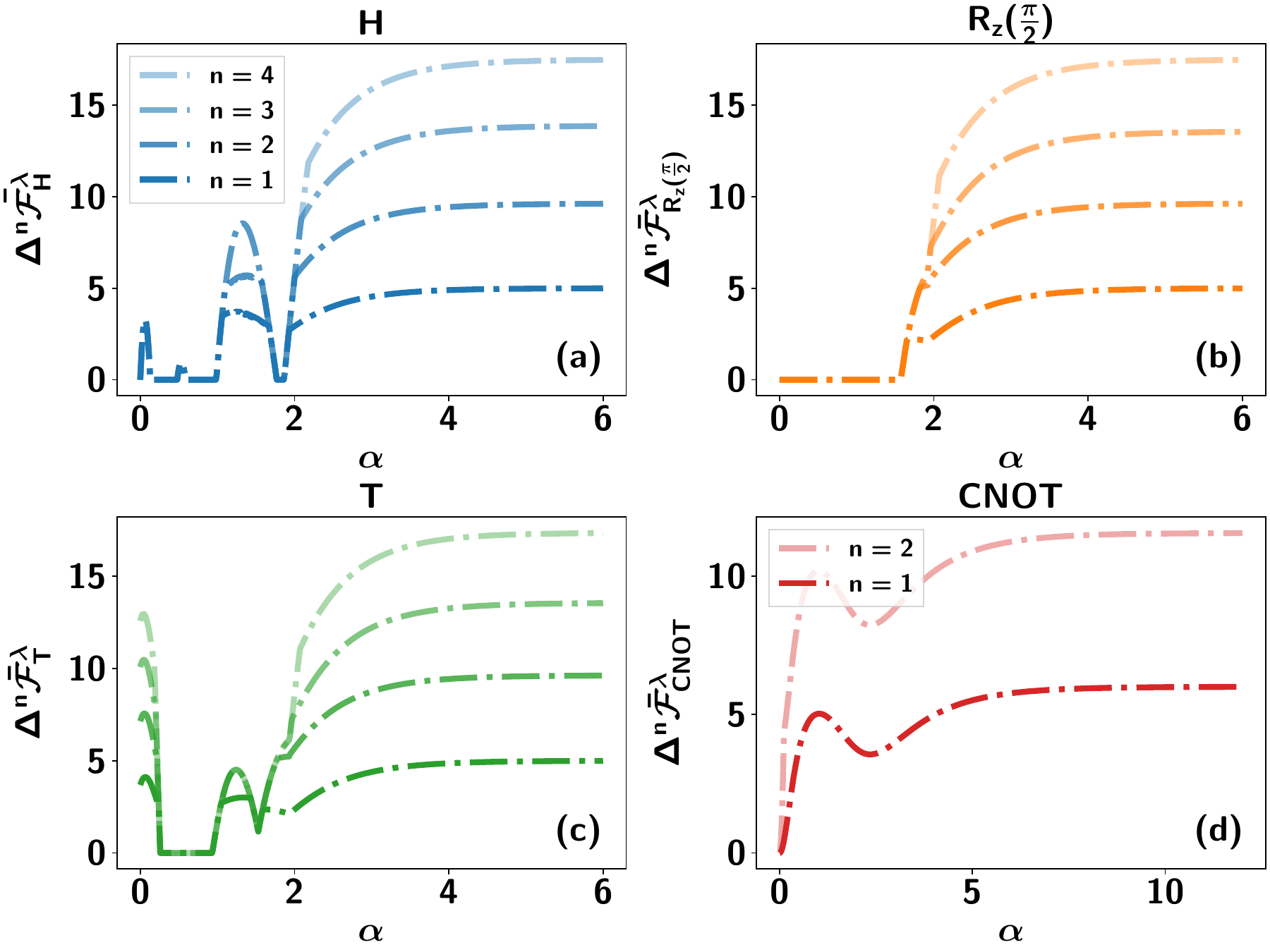}
    \caption{(Color online.) Robustness of fidelity (RF), \(\Delta^n\bar{\mathcal{F}}_{\mathcal{G}}^{\lambda}\) (y-axis) with respect to the fall-off rate $\alpha$ (x-axis) for a fixed value of unsharp parameter \(\lambda=0.85\) involved in measurements. Fidelities are plotted for (a) Hadamard, (b) \(\frac{\pi}{2}\)-phase, (c) \(T\), and (d) $CNOT$ gates. Dark to light sequential shades of a particular plot imply an increasing number \(n\) of unsharp measurements. Note that $\Delta^n\bar{\mathcal{F}}_{\mathcal{G}}^{0.85}$ saturates to approximately $5\%,9\%,13\%$, and $17\%$ for $n=1,2,3$, and $4$, respectively, for all the single qubit gates. For $CNOT$ gates,   $\Delta^n\bar{\mathcal{F}}_{{CNOT}}^{0.85}$  reaches maximum of $6\%$ and $11\%$ approximately, with $n=1$ and $2$ respectively. Both axes are dimensionless. } 
    \label{fig:unsharp_delta_H}
\end{figure}

We now move to the scenario in which noisy measurements are carried out on the first $n( > 1$) number of qubits of $V_m$. 
In particular, for a fixed value of $\alpha$, as the number of unsharp measurements increases, i.e., $n$ increases,  $\nonumber\Delta^n\bar{\mathcal{F}}_{\mathcal{G}}^{\lambda}$ increases and saturates to a maximum value for $\alpha\gtrsim 5$ as demonstrated in Fig.~\ref{fig:unsharp_delta_H}. It clearly demonstrates that the increase of the number of unsharp measurements implies more entanglement among qubits in $,V_m$ which diminishes the fidelity at $V_O$ since the trade-off between the entanglement-creation via the evolving Hamiltonian and -destruction due to measurements is responsible for the successful implementation of MBQC. Precisely, for the single-qubit gates, $\Delta^n\bar{\mathcal{F}}_{\mathcal{G}}^{0.85}$ reaches their maximum errors of approximately $5\%,9\%,13\%$ and $,17\%$ for $n=1,2,3$, and $4$, respectively, independent of $\mathcal{G}$. On the other hand, $\Delta^n\bar{\mathcal{F}}_{{CNOT}}^{0.85}$ is minimum at $\alpha\approx 0$ and reaches maximum of $6\%$ and $11\%$ approximately, for $n=1$ and $2$ respectively. Hence, the adverse effects of unsharp measurements are more pronounced when $\alpha$ is high, close to the NN model, compared to the region with $\alpha\lesssim3$. 


\section{Impact of disorder on MBQC}
\label{sec:disorder}

The protocol outlined above operates optimally when the coupling parameter \( J \) is fixed for all sites, representing an ideal scenario. In a practical situation, deviations are inevitable, and hence the coupling between the sites fluctuates. It is tempting to predict that this site-dependent \( \{J_i\} \) may decrease the performance, although there are several counterexamples in literature where disorder either shows some increment or robustness \cite{ahufinger06, Prabhu_2011, Shtanko_2020, ghosh_2020, konar_2022,konar_2022b, halder_2025}. Specifically, we analyze the average fidelity of the implemented gates in the presence of disorder in \( \{J_i\} \), wherein the timescale associated with fluctuations in \( \{J_i\} \)s is assumed to be significantly longer than the characteristic evolution timescale of the quantum system and hence quenched averaging can be carried out. 

We are here interested to compute the quenched average fidelity for a given gate $\mathcal{G}$ by performing the ensemble-averaged quantities with integration over the probability distribution of \( \{J_i\} \). Accordingly, for a physical quantity \( \mathcal{Q} \), the quenched-average value, denoted by \( \langle \mathcal{Q} \rangle \), is given by
\begin{equation}
\langle \mathcal{Q} \rangle^\sigma = \int \mathcal{Q}(J) P(J) \, dJ, \label{eq:dis}
\end{equation}
where \( P(J) \) is a probability distribution with mean \( \bar J \) and standard deviation (SD) \( \sigma\). We refer to this \(\sigma\) as the disorder strength, with the limiting case, \( \sigma = 0 \), recovering the ordered scenario discussed previously. For our analysis, we sample values of \( \{J_i\} \) from a Gaussian distribution having unit mean and varying SD. In particular, the investigation of the quenched-average gate fidelity, \(\langle \Bar{\mathcal{F}}_{\mathcal{G}}\rangle^\sigma\) is obtained by evaluating the integral numerically with mean $\bar J=1$, for various $\sigma$s and for \(10^3\) realizations, thereby revealing the impact of disorder on the overall performance of the protocol. Moreover, the same optimal corrective unitaries are used as those calculated with $J=\bar J =1$ for calculating \(\langle \Bar{\mathcal{F}}_{\mathcal{G}}\rangle^\sigma\). As mentioned before, this assumption is reasonable since disorder appears in the system during implementation. 

\begin{figure}
    \centering
    \includegraphics[width=\linewidth]{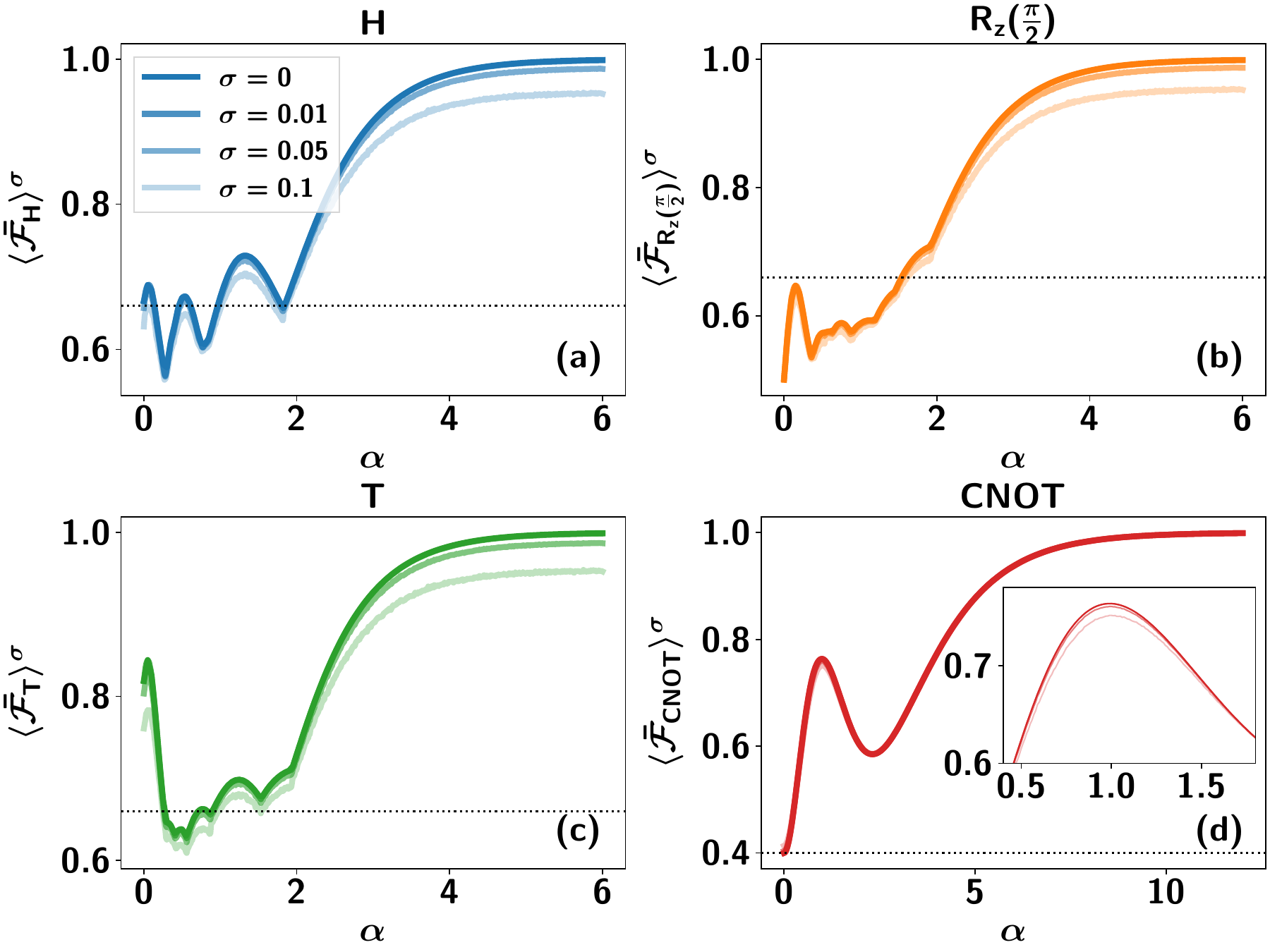}
    \caption{(Color online.) The quenched-averaged gate fidelity, \(\langle \Bar{\mathcal{F}}_{\mathcal{G}}\rangle^\sigma\) (abscissa) against fall-off rate $\alpha$ (ordinate) for various quantum gates. 
    Dark to light shades here signify the increasing value of the disorder strength, $\sigma$. The black dotted line represents $F_c$ for the corresponding gates. All other specifications are the same as in Fig. \ref{fig:unsharp_delta_H}. Inset in (d) indicates the gate fidelity in the presence of disorder with different \(\sigma\)s when \(\alpha \in (0.4, 1.8)\). Both axes are dimensionless.} 
    \label{fig:disorder}
\end{figure}

{\bf Effect on single-qubit gates:} For arbitrary interaction strength $\alpha$, with increasing $\sigma$, $\langle \Bar{\mathcal{F}}_{\mathcal{G}} \rangle$ decreases monotonically (see Fig.~\ref{fig:disorder}). Although the decrease of \(\langle \Bar{\mathcal{F}}_{\mathcal{G}}\rangle^\sigma\) is not massive, possibly indicating its robustness against impurities. Still, to quantify the detrimental effect of disorder in $J$, we define the relative error as $\delta_\sigma^{\mathcal G}= ((\max(\bar{\mathcal{F}}_{\mathcal{G}},2/3) -\max(\langle \Bar{\mathcal{F}}_{\mathcal{G}} \rangle^\sigma,2/3))/\bar{\mathcal{F}}_{\mathcal{G}})\times 100\%$. Similar to the effect of noisy measurements, we again find that the impact of disorder is more apparent for high $\alpha$ ($\alpha\gtrsim4$) than that for low $\alpha$ values ($\alpha\approx2$ and below). Specifically, for $\sigma=0.01$, and at \(\alpha_s\) (where the saturation of the fidelity occurs), the relative error for any single qubit gate, i.e., $\delta_{0.01}^{\mathcal G}$ has a maximum value of approximately $0.04\%$ with $\mathcal G=\{H,R_z(\pi/2),T\}$ while as the standard deviation increases, e.g., for $\sigma=0.05$ and $0.1$, $\delta_{\sigma}^{\mathcal G}$ increases to $1.15\%$ and $4.60\%$, respectively. All these observations confirm that the gate fidelities are resilient against the fluctuations in the evolution operator and measurements.

{\bf Impact of disorder on $CNOT$ gate:} Again, disorder in $J$ does not have any significant negative impact on fidelity of the $CNOT$ gate.  For  small $\sigma$ values, $\langle \Bar{\mathcal{F}}_{CNOT} \rangle^\sigma\approx\Bar{\mathcal{F}}_{CNOT}$  $\forall \alpha$ as evident from Fig.~\ref{fig:disorder}(d). For large values of \(\sigma\), say \(\sigma=0.1\), \(\delta_{0.1}^{\text{CNOT}}\) has some fluctuations on the both side of \(\alpha\approx2\),  and approximately \(2\%\) of deviation from the ordered case can be observed ( see inset of Fig.~\ref{fig:disorder}(d)). \(\delta_{0.1}^{\text{CNOT}}\)is negligible for \(\alpha\gtrsim3\), \(\mathcal{O}(10^{-2})\), which saturate to \(\mathcal{O}(10^{-3})\) even before \(\alpha_s\).


\section{Conclusion}
\label{sec:conclu}

Long-range (LR) interacting spin systems, where interaction strengths decay with the distance between spins, naturally arise in platforms such as trapped ions and cold atoms in optical lattices. When initialized in a product state, evolution under LR Hamiltonians can generate highly entangled cluster-like states, referred to as weighted graph states (WGS), that exhibit genuine multipartite entanglement comparable to that of ideal cluster states. This makes LR systems promising candidates for implementing measurement-based quantum computation (MBQC). However, successful realization of quantum gates in MBQC depends not only on the quality of entanglement but also on other crucial factors, such as the choice of projective measurements, appropriate corrective unitaries, and precise control over system parameters.

Within the MBQC framework, we employed states generated by LR interactions to implement quantum gates and analyzed how the fall-off rate of interaction, governed by a power-law, affects gate performance. Our findings revealed that the average gate fidelity, a key indicator of implementation accuracy, is strongly influenced by this fall-off rate and demonstrated conditions under which the fidelity approaches unity, even in systems that are not strictly produced via nearest-neighbor (NN) models. For both single- and two-qubit gates,  we showed that the fidelity optimization requires adjusting the corrective unitaries based on interaction range. Furthermore, we examined the effects of unsharp measurements, where residual entanglement persists, and of random disorder in interaction strengths. In both cases, we observe that the gate fidelities remain robust. These results emphasize the practical viability of WGS as a resource for MBQC under realistic experimental conditions.

 \acknowledgements
 We acknowledge the use of \href{https://github.com/titaschanda/QIClib}{QIClib} -- a modern C++ library for general-purpose quantum information processing and quantum computing (\url{https://titaschanda.github.io/QIClib}) and cluster computing facility at Harish-Chandra Research Institute. PH acknowledges ``INFOSYS
Scholarship for senior students".

\appendix
\section{Cluster states}
\label{app:cluster}
Qubit cluster states are defined on a cluster $\mathcal C$, a connected subset of a simple cubic lattice $\mathbb Z^d$ 
with $N$ lattice points where $d\geq 1$ is the dimension of the lattice. The lattice points of the cluster $\mathcal C$ forms a set of vertices $V=\{i|i\in\mathcal C\}$ (replaced by qubits) connected to nearest-neighbours via edges which form  the set $E=\{(i,j)|i,j\in \mathcal C, j\in \text{nghb}(i)\}$. Here, $\text{nghb}(i)$ is the set of all nearest-neighbouring sites of $i$. The cluster states $\ket {\psi_{\mathbf k}}_{\mathcal C}$ corresponding to the cluster (or graph) \footnote{We use the words ``graph'' and ``cluster'' alternatively, since cluster states are a special kind of graph states with only nearest-neighbour interactions.} $\mathcal C(V,E)$ satisfies $K^{(i)}\ket {\psi_{\mathbf k}}_{\mathcal C} = (-1)^{k_i}\ket{\psi_{\mathbf k}}_{\mathcal C}$ $\forall i\in V$. Here, the string $\mathbf k=k_1k_2\ldots k_N$ with $\{k_i\in\{0,1\}|i\in \mathcal C\}$. $K^{(i)}$s are known as the stabilizers of the cluster state when $k^{(i)}=0$ $\forall i$ \cite{rauss_pra2003}. We denote such a cluster state as $\ket{\psi}_{\mathcal C}$. 

Physical realization of the cluster state in the laboratory requires initializing the qubits of the $\mathcal C$ in the eigenstate of $\sigma_x$ with $+1$ eigenvalue, i.e., $\ket +=\frac{\ket{0}+\ket{1}}{\sqrt{2}}$. Here \(\ket{0}\) and \(\ket{1}\) are the eigenstate of \(\sigma_z\) with $+1$ and $-1$ respectively. This serves the purpose of generating cluster states, $\ket{\psi}_{\mathcal C}$ at times $t=n\pi$ $\forall n\in\mathbb Z^+$ by evolving  the initial state $\otimes_{i\in V}\ket +_i$  with $e^{-\iota H_{NN}t}$ where $\iota=\sqrt{-1}$.

\section{Average gate fidelity of $T$-gate with restricted set of corrective unitary}
\label{app:tgate}

The average gate fidelity of $T$-gate using  $U_{c,T}^{\infty,\mathbf{s}}=\sigma_x^{s_2+s_4}\sigma_z^{s_1+s_3}$ as corrective unitary is calculated as
\begin{eqnarray}
\nonumber   \bar{\mathcal{F}}_{T}^r= \frac{1}{96} &\Big(& 53 + \cos2a_2 +8\cos a_2 + 6\cos a_3\\ \nonumber
&+& 4\cos b_3 + 2\cos(b_3+a_2) + 2\cos(b_3+a_3) \\ \nonumber
&+&\cos a_2\cdot[4\cos a_3 + \cos a_4 + 3\cos(a_2+a_4) \\ \nonumber
&+& 2\cos(a_3+a_4)+ 6\cos b_4 + \cos(a_3+b_4) \\ \nonumber
&+& 3\cos(b_3+b_4)-\sin a_4 + \sin(a_2+a_4) \\ \nonumber
&+& 2\sin(a_3+a_4)-2\sin b_4-\sin(a_3+b_4) \\
&+& \sin(b_3+b_4)]\Big),
\label{eqn:fid_T}
\end{eqnarray}
where $a_j=j^{-\alpha}\pi$ for $j=2,3,4$, $b_3=a_2+a_3$ and $b_4=b_3+a_4$ are defined as in the main text.


\bibliography{bib.bib}
\end{document}